\documentstyle[epsfig,sprocl]{article}

\def\Journal#1#2#3#4{{#1} {\bf #2}, #3 (#4)}

\def\be{\begin{equation}}
\def\ee{\end{equation}}
\def\bea{\begin{eqnarray}}
\def\eea{\end{eqnarray}}

\begin{document}

\title{HIGH RESOLUTION PIXEL DETECTORS FOR $e^{+}e^{-}$ LINEAR COLLIDERS}

\author{ M.CACCIA\footnote{Corresponding Author; e-mail: Massimo.Caccia@mi.infn.it},
 R.CAMPAGNOLO, C.MERONI }

\address{Dipartimento di Fisica, Universita' di Milano and I.N.F.N,\\
Via Celoria 16, 20133 Milano, Italy}

\author{ W.KUCEWICZ, G.DEPTUCH }

\address{Department of Electronics, University of Mining and Metallurgy,\\
Mickiewicza 30, Kracow, Poland}

\author{ A.ZALEWSKA}

\address{Institute of Nuclear Physics,\\
ul. Radzikowskiego 152, Krakow, Poland}

\author{ M.BATTAGLIA, K.OSTERBERG, R.ORAVA}

\address{Helsinki Institute of Physics,\\
P.O.Box 9, FIN-00014 University of Helsinki, Finland}

\author{ S.HIGUERET, M.WINTER}

\address{Institut de Recherches Subatomique,\\
23, rue du Loess BP28, F-67037- Strasbourg cedex 2, France}

\author{ R.TURCHETTA }

\address{Laboratoire d'Electronique et de Physique des Systèmes Instrumentaux,\\
23, Rue du Loess BP20, F-67037- Strasbourg cedex 2, France}

\author{P.GRABIEC, B.JAROSZEWICZ, J.MARCZEWSKI}

\address{Institute of Electron Technology,\\
al. Lotnikov, 32/46, 02-468 Warszawa, Poland}

\maketitle\abstracts{
The physics goals at the future $e^{+}e^{-}$ linear colliders require
high performance vertexing and impact parameter resolution. Two possible
technologies for the vertex detector of an experimental apparatus are outlined in
the paper: an evolution of the Hybrid Pixel Sensors 
already used in high energy physics experiments and  
a new detector concept based on the
monolithic CMOS Sensors. }
 
\section{Introduction}
Precision measurements of Top quark and Higgs boson physics are in the reach of the 
next generation of   $e^{+}e^{-}$ linear colliders, operating at a centre of mass energy
ranging from the $ Z^o$ pole to ${\rm 1~TeV}$. 
High granularity, impact parameter resolution, secondary vertex
reconstruction and jet flavour tagging are the 
essential tools for these measurements and define the figures of merit for a Vertex Detector.\\

In the 1996 Joint DESY/ECFA study, the minimal requests on the impact parameter
resolutions in the ${\rm R\Phi}$ plane and along the beam direction were defined to be \cite{ecfa}
${\rm \delta(IP_{R\Phi})~=~10~\mu m \oplus \frac{30~\mu m~GeV/c}{p~sin^{3/2}(\theta)} }$ and
${\rm \delta(IP_{z})~=~20~\mu m \oplus \frac{30~\mu m~GeV/c}{p~sin^{5/2}(\theta)} }$
with a total material budget below ${\rm 3\%~X_{0}}$ for the complete Vertex Detector, fitting 
between the beam pipe originally at a ${\rm 2~cm}$ radius and the intermediate tracker starting at 
${\rm 12~cm}$. Conceptual design of vertex trackers based on hybrid pixel sensors \cite{hpd} and CCD's
\cite{chris}
were proposed and Research \& Development plans defined. 

Since then, several analysis remarked the advantages of possibly improved performances,
both in terms of asymptotic resolution and multiple scattering. 
This triggered the quest for lightweight technologies that could provide space point
informations along the particle track with at least ${\rm 10~\mu m}$ resolution. At the same time,
an improved final focusing scheme allowed to shrink the beam pipe to ${\rm 1~cm}$ radius, with the
inner sensitive layer at ${\rm ~1.2~cm}$.

In the following, two possible  detector technologies are presented: the former is based
on the development of the Hybrid Pixel Detectors  used in DELPHI \cite{df} and WA-97
\cite{wa}
and being finalized for the LHC experiments \cite{atlas,cms,alice};
 the latter is based on the evolution of 
monolithic CMOS imagers \cite{cmos1}
to achieve the sensitivity to minimum ionizing particles.   
 
\section{Hybrid Pixel Detectors}
\subsection{General concepts}\label{subsec:HPD}
The achievement of a ${\rm 10~\mu m}$ resolution in a Silicon detector can be accomplished
probing the diffusion of the charge carriers locally generated around the impinging particle
trail. Given the diffusion characteristics, this require a microstrip or pixel pitch 
well below ${\rm 50~\mu m}$ and an analog output to interpolate the signals on neighbouring cells.

While this is feasible with the 1d microstrip detectors \cite{anna}, the need to 
integrate the front-end
electronics in a cell matching the detector pattern defines the ultimate pitch
in 2d pixel detectors. At the moment,
the most advanced read-out chips have a minimum cell dimension of 
${\rm 50 \times 300~\mu m^{2}}$, produced in 0.8 CMOS technology, implemented in a 
rad-hard process. On one hand, the trend of the VLSI development and the recent studies
\cite{snow}
 on intrinsic
radiation hardness of deep submicron CMOS technology certainly allow to assume a relevant 
reduction in the cell dimensions on a mid-term. On the other hand, a detector design overcoming 
this basic limitation is worth being considered.

\begin{figure}
\begin{center}
\epsfig{file=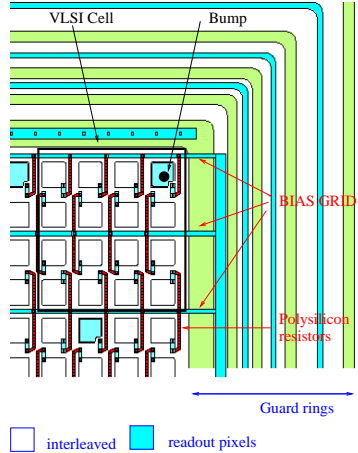,bbllx=0,bblly=0,bburx=465,bbury=600,height=6cm}
\caption{Generic layout of the proposed detector, corresponding to a ${\rm 50~\mu m}$ pitch
in both dimensions; the other structures differ by the pitch only}
\label{fig:corner}
\end{center}
\end{figure}

What is being proposed is a layout inherited from the microstrip detectors 
\cite{hyams} where it is assumed 
to have a readout pitch n times larger than the pixel pitch (see for instance fig. 
~\ref{fig:corner} for ${\rm n=4}$).
In such a configuration, the charge carriers created underneath an interleaved pixel will induce
a signal on the output nodes, capacitively coupled to the interleaved pixel.
In a simplified model where the detector is reduced to a capacitive network, 
the ratio of the signal amplitudes on the output nodes at the left hand 
side and right hand side of the interleaved pixel 
(in both dimensions) should have a linear dependence on the particle position.
The ratio between the inter-pixel capacitance and the
pixel capacitance to backplane plays a crucial rule in the detector design, 
as it defines the signal amplitude reduction
(an effective charge loss) at the output nodes and at last the sustainable number of 
interleaved pixels. 
Recent results \cite{krammer} on ${\rm 200~\mu m}$ readout pitch microstrip detectors have been published, and a 
${\rm 10~\mu m }$ resolution has been achieved in a layout
with 3 interleaved strips (${\rm 50~\mu m}$ strip pitch) and for a ${\rm S/N \approx 80}$.
 Similar results may be expected in a pixel
detector, taking into account a lower noise is achievable because of the intrisically smaller load
capacitance and the charge is possibly shared on four output nodes reconstructing the particle
position in two dimensions. Improvements are certainly possible sampling the diffusion with 
a smaller pitch.

\subsection{Detector prototypes and electrostatics characterization}\label{subsec:depro}
Prototypes of detectors with interleaved pixels have been designed in 1998 and delivered in january
1999.

The layout of one of the structures is shown in fig.~\ref{fig:corner}.
A series of guard rings defines the
detector sensitive area. A bias grid allows the polarization of the interleaved pixels too; each
${\rm p^{+}}$ implant is connected to the metal bias line by polysilicon resistors in the  ${\rm 1-3~M
\Omega}$ range.
A metal layer is deposited on top of the pixels to be connected to the VLSI cell. The
backplane has a meshed metal layer to allow the use of an IR diode for charge collection studies.
In a 4" wafer 36 structures were fit, for 17 different layouts; a VLSI cell of ${\rm 200 \times 200~\mu
m^{2}}$ or ${\rm 300 \times 300~\mu m^{2}}$ was assumed and detectors with a number of interleaved pixels
ranging between 0 and 3 and different areas were designed. 

Ten high resistivity wafers ${\rm (5-8~k\Omega cm)}$ were processed 
\footnote{at the Institute of Electron Technology, Warszawa, Poland} together with an equal number of low
resistivity wafers for process control, the details of which have been outlined in \cite{az}.
Two wafers were retained by the factory for a destructive analysis and two others were stored 
for later use.
All of the structures on five undiced wafers were visually inspected, tested up to ${\rm 250~V}$ and 
characteristics I-V and C-V curves produced; the results may be summarized as follows:

$\bullet$ two wafers suffered from processing problems. As a consequence of it, one wafer had 
a high leakage current (${\rm \geq 1 \mu A}$) even at very low voltages on most of the structures;
the second wafer had interrupted metal lines, making the bias grid inefficient. The former problem
is possibly connected to the Al pattern plasma etching; the latter to a not optimal planarization 
of the device.

\begin{figure}
\begin{center}
\epsfig{file=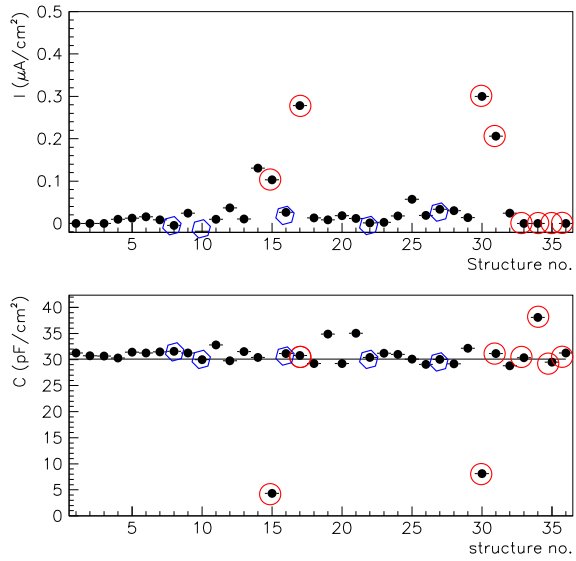,bbllx=0,bblly=7,bburx=500,bbury=516,height=8cm}
\caption{Leakage current and capacitance at full depletion voltage for all of the
36 structures in one of the good wafers; circles define the detectors rejected
because of breakdowns below ${\rm 100~V}$; hexagons identify rejected detectors because
of interrupted metal lines. The line corresponds to the expected capacitance per unit area.}
\label{fig:waf05}
\end{center}
\end{figure}

\begin{figure}
\begin{center}
\epsfig{file=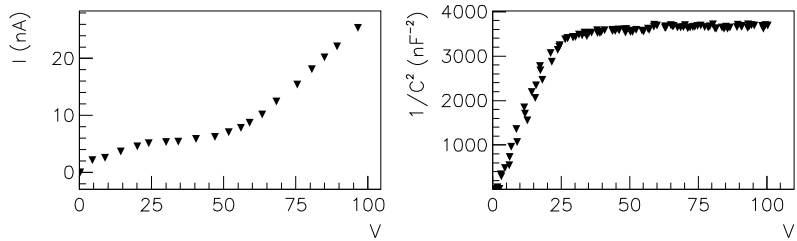,bbllx=3,bblly=355,bburx=528,bbury=528,width=11cm}
\caption{${\rm I~vs.~V}$ and ${\rm 1/C^{2}~vs.~V}$ curves for a typical good detector}
\label{fig:ivcv}
\end{center}
\end{figure}

$\bullet$ three wafers had extremely good characteristics, with a mean current \linebreak
${\rm \approx 50~nA/cm^2}$
at full depletion. Structures were classified as good detectors if no breakdown was observed
below ${\rm 100~V}$, the leakage current at depletion voltage was below ${\rm 1~\mu A}$ with a smooth trend
vs. the applied voltage and  no faults in the line pattern were
detected by visual inspection.
According to these criteria, 55/89 structures were accepted.
In fig.~\ref{fig:waf05} the value of the currents and detector capacitances at
 depletion voltage are shown for all of
structures in one of the three wafers. In fig.~\ref{fig:ivcv}
the typical ${\rm I~vs.~V}$ and ${\rm 1/C^{2}~vs.~V}$ curves are shown.

While the ${\rm C~vs.~V}$ curves behave as expected, the current has a peculiar trend. After a plateau is 
reached at full depletion, the current takes off at values in the ${\rm 50-70~V}$ range. For most of the
structures it is a mild increase, but most of the rejected detectors are characterized by a 
steep slope, eventually ending with a breakdown below ${\rm 100~V}$. Independent measurements of the 
guard ring and bias grid current has shown the latter is responsible for the increase, that might
be connected to sharp edges where the electric field achieves high values. A full device simulation
is planned to help understanding this feature.

\subsection{Outlook}
The prototypes have not shown any design fault even if processing and layout optimization 
has to be considered. On a short term, measurements of the inter-pixel and backplane capacitances
are planned, completing the electrostatics device characterization. 
A charge collection study will follow, relying on a low noise strip detector analog chip and
an IR light spot shone on the meshed backplane. These measurements will be a  proof of
principle of the proposed device and define the fundamentals for a further iteration, aiming at
a ${\rm 25~\mu m}$ pitch.

The device thickness is a particularly relevant issue for the application of Hybrid Pixel Sensors
in a linear collider experiment. The minimal thickness is defined by both the detector performances
and the backthinning technology for bump bonded assembly.
Industrial standards guarantee backthinning down to ${\rm 50~\mu m}$ and a procedure to obtain thin
Hybrid Pixel detectors is being tested \cite{sott}. The small load capacitance of the pixel cells shall
guarantee an extremely high ${\rm S/N}$. Scaling what was obtained for microstrip detectors, the desired
resolutions might be obtained with a ${\rm 200~\mu m}$ thick detector 
(a ${\rm 250~\mu m}$ (${\rm 0.27\% X_o }$) thick assembled
device). Moreover, the mechanical structure of a three layer vertex detector based on Hybrid Pixel
Sensors is being designed and a realistic material budget evaluated.

\section{Monolithic Pixel Sensors}

\subsection{General concepts}
\begin{figure}
\begin{center}
\epsfig{file=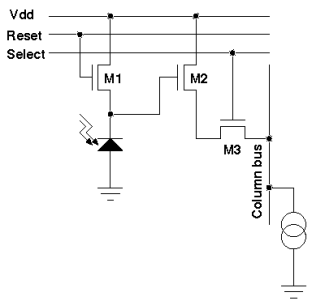,bbllx=114,bblly=208,bburx=496,bbury=580,height=4cm}
\caption{The baseline architecture of a monolithic APS. 
Transistor M1 resets the photosite to reverse 
bias; transistor M3 is a column select switch, 
while transistor M2 is the input of a source follower, whose 
current source is located outside the pixel}
\label{fig:cmos1}
\end{center}
\end{figure}

In the early 90's monolithic pixel sensors have been proposed as a viable 
alternative to CCD's in visible imaging \cite{cmos1}. These sensors are made in a standard 
VLSI technology, often CMOS, so they are usually called 
CMOS imagers. Three main architectures have been proposed, namely: Passive 
Pixel Sensors (PPS) and Active Pixel Sensors (APS) with photodiode or 
photogate. 
In the former, a photodiode is integrated in each pixel together with a 
selection switch, while in the latter  (fig.~\ref{fig:cmos1}), 
three transistors are usually 
integrated together with a photosite. 

Today, most of the sensors are based on APS because of their superior noise 
performances: 
electron noise can be as low as ${\rm 4.5~e^{-}}$ r.m.s. at room temperature.

The use of standard CMOS technology gives APS several advantages with respect to the 
more generally used CCD's: they are low cost; inherently radiation hard; several functionalities can be 
integrated on the sensor substrate, including random access; they consume very little power as
the circuitry in each pixel 
is active only during the readout and there is no clock signal driving large capacitances.
 

%
%
%
%
%
%
%

Because of these characteristics, CMOS sensors are the favoured technology for 
demanding applications which are typically 
found in space science.

\subsection{CMOS sensors for charged particle detection}

\begin{figure}
\begin{center}
\epsfig{file=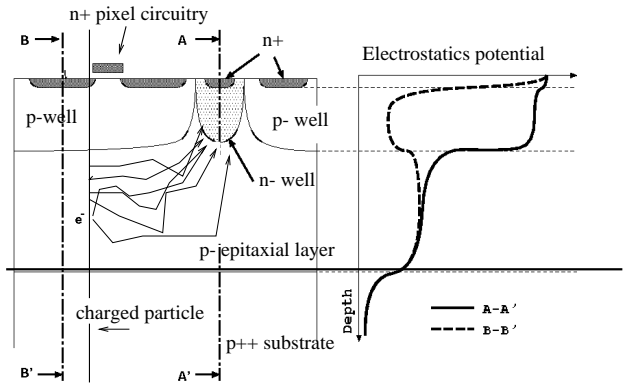,bbllx=0,bblly=0,bburx=538,bbury=320,height=5cm}
\caption{The proposed pixel structure. 
The circuitry is integrated in the p-well while the photosite is an 
n-well diode on the p-epitaxial layer. 
Because of the difference in doping level, the p-well and the p++ 
substrate act as reflective barriers. 
The generated electrons are collected in the n-well.}
\label{fig:cmos2}
\end{center}
\end{figure}

In visible light applications, special care is taken to maximise 
the fill factor, i.e. the 
fraction of the pixel area that is sensitive to the light. 
Because of the transistors, fill 
factors in CMOS sensors are relatively low (in the order of 30\%). 
This can be a severe 
limitation in high-energy physics application, 
if no special care is taken. One of us \cite{turch} 
proposed to integrate a sensor 
in a twin-well technology with an n-well/p-substrate 
diode in order to 
achieve 100\% fill factor for ionizing particle detection (fig.~\ref{fig:cmos2}). 
This technique 
has already proven its 
effectiveness in visible light applications \cite{cmos2}
reducing the blind area to 
the metal lines, opaque
for visible light but not for charged particles.

CMOS sensors could achieve high spatial resolution: the pixel size is usually 
between 10 and 20 times the Minimal 
Size Feature of the used technology, which means that ${\rm 10~\mu m}$ pitch is possible, 
and hence spatial resolution better than ${\rm 3~\mu m}$ even with a binary readout.
At the same time, very low multiple scattering is introduced as the substrate can in principle 
be thinned down to a few microns.
A charge particle CMOS detector would also benefit from the generic characteristics of these devices,
including low power dissipation, radiation resistance of deep submicron CMOS technologies 
and low cost.  

%
%
%
%
%
%
%

\subsection{CMOS sensors for a Linear Collider}

In order to prove the effectiveness of CMOS sensors 
for the Next Linear Collider, an 
R\&D program has been initiated by the Strasbourg group. 
Some existing commercial 
devices are currently under test. Since the performances 
of the sensors depend on details of the fabrication process, 
a full-custom design of a first 
prototype sensor (MIMOSA = MIP MOS APS) has been done in a ${\rm 0.6~\mu m}$ CMOS 
technology. The circuit will be back from the foundry in autumn 1999.

A preliminary detector design of a Microvertex Detector based on CMOS sensors was also
completed. 
The detector is supposed to be made of 5 layers, the 
inner most having a radius of 1.2 cm. It is made of a cylindrical barrel part 
${\rm (|cos(\theta)|<0.90)}$ associated to forward and backward conical and disk-like 
extensions ${\rm (0.90<|cos(\theta)|<0.99)}$, 
intercepting with at least 4 layers charged 
particles produced at polar angles ranging from 6 to 174 degrees. 
The sensors are 
assumed to be ${\rm 50~\mu m}$ thick squares of ${\rm 1.4 \times 1.4~ cm^2}$ area, 
with an active surface close to 80\%. 
Assuming a few per-cent overlap between the active 
surfaces of neighbour sensors, 
about 5500 units are needed to cover the ${\rm 1.7~m^2}$ area of the detector.
Since the sensors have low power dissipation, 
a mechanical support made of ${\rm 100~\mu m}$ 
thick, ${\rm 7~mm}$ large, thermal diamond rods was considered. 
A detailed simulation 
showed that such a device, connected to a light system of thin 
cooling pipes, would 
provide enough thermal conduction by itself to evacuate the 
heat from the sensors, thus
 substantially reducing the material seen by the particles. 
The simulation of the 
mechanical constraints showed that the bending of the rods 
should nowhere exceed 
a ${\rm 30~\mu m}$ sagitta.
As a further advantage, diamond aluminised 
with a few micron thick 
layer could fan in/out all the electrical signals. 
Globally, the material budget is such that 
particles crossing the 5 detector planes would in average see a 
total amount of 0.8\% 
radiation length in the barrel and 2.9\% in the forward-backward parts,
values making the CMOS sensor based vertex detector very competitive.

\section{Conclusions}
Two detector technologies suitable for a Vertex Detector at the next
generation of ${\rm e^{+}e^{-}}$ colliders have been presented in the paper.
Hybrid Pixel Sensors could achieve the desired performances overcoming the 
limitations defined by the electronics cell dimension, mating the pixel, and developing
a dedicated analog readout chip. 
Charge sharing and S/N ratio are the critical issues for these
detectors as they determine both the resolution and the minimal detector thickness.
Detector prototypes have been produced; the electrostatics characterization
is ongoing and the first results are positive.
Monolithic CMOS sensors could achieve an excellent resolution introducing a very low
multiple scattering. Impressive results as visible light detectors have been
recently obtained and a custom designed device optimized for ionizing particles
have been submitted.

\end{document}